\documentstyle[12pt,psfig]{article}

\title{{\bf Nuclear medium modifications of the NN interaction  via quasielastic ($\vec p,\vec p\ '$) and ($\vec{p},\vec{n}$) scattering}
\footnote{submitted to Phys.\ Rev.\ C; manuscript CD6345}}
\author{G. C. Hillhouse, B. I. S van der Ventel, S. M. Wyngaardt and
P. R. de Kock\\ {\normalsize\it Department of Physics, University of Stellenbosch, Stellenbosch 7600, South Africa }}
\begin{document}
\maketitle
\begin{abstract}
Within the relativistic PWIA, spin observables have been recalculated 
for quasielastic ($\vec p,\vec p\ '$) and ($\vec p,\vec n$) reactions on a $^{40}$Ca target.  The 
incident proton energy ranges from 135 to 300 MeV while the transferred 
momentum is kept fixed at 1.97~fm$^{-1}$.  In the present calculations, 
new Horowitz--Love--Franey relativistic NN amplitudes have been 
generated in order to yield improved and more quantitative 
spin observable values than before. The sensitivities of the various 
spin observables to the NN interaction parameters, such as 
(1) the presence of the surrounding nuclear medium, (2) a pseudoscalar 
versus a pseudovector interaction term, and (3) exchange effects, point 
to spin observables which should preferably be measured 
at certain laboratory proton energies, in order to test current 
nuclear models. This study also shows that nuclear medium effects 
become more important at lower proton energies ($\leq$ 200 MeV).  
A comparison to the
limited available data indicates that the relativistic 
parametrization of the NN scattering amplitudes in terms of only
the five Fermi invariants (the SVPAT form) is questionable.
\end{abstract}
\newpage
Considerable attention has been devoted to the
measurement and interpretation of inclusive ($\vec p,\vec p\ '$)
and ($\vec p,\vec n$) polarization observables at the quasielastic peak \cite{Te88,Te91,Nu94}.
At moderate momentum transfers (1 $\leq$ q $\leq$ 2 fm$^{-1}$)
quasielastic scattering becomes the dominant mechanism for 
nuclear excitation. It is considered to be a single--step process 
whereby a projectile particle knocks out a single bound nucleon in 
a target nucleus while the remainder of the nucleons remain inert. 
This process is characterized by a broad bump in the excitation
spectrum, the centroid of which nearly corresponds to free NN kinematics,
and a width resulting from the initial momentum distribution of the
struck nucleon. At the momentum transfers of interest, shell effects are unimportant \cite{Ch89}
and the quasielastic peak is well separated from discrete states in the excitation spectrum.
Hence, deviations of the polarization transfer observables from the corresponding free NN values could be 
attributed to medium modifications of the free NN interaction.

The failure of all nonrelativistic Schr\"{o}dinger--based models \cite{Te88} 
to describe the quasielastic 
($\vec p,\vec p\ '$) analyzing power at 500 MeV lead to the development of the Relativistic 
(Dirac) Plane Wave Impulse Approximation (RPWIA) \cite{Ho86a,Ho86b,Ho88}, where the NN amplitudes are based 
on the Lorentz invariant parametrization of the standard five Fermi invariants 
(the so--called SVPAT form) and the target nucleus is treated as a Fermi gas.
Medium effects (often referred to as relativistic effects) are incorporated by 
replacing free nucleon masses in the Dirac plane waves with effective projectile and target 
nucleon masses in the context of the Walecka model \cite{Se86}. 

Despite the successful RPWIA prediction of A$_{y}$ \cite{Ho88}, most of the other five independent spin transfer observables allowed by 
parity and time-reversal invariance, $\mbox{D}_{n n}, \mbox{D}_{s' s}, \mbox{D}_{{\ell}' \ell}, 
\mbox{D}_{s' \ell}$ and D$_{{\ell}' s}$, seem to favor relativistic calculations based on free
nucleon masses. However, the original RPWIA predictions \cite{Ho86a,Ho86b,Ho88,Ha88} were based on crude 
assumptions and unrefined input. For example, 
a 10\% uncertainty in effective mass values translates into 30\% effects on some 
polarization transfer observables \cite{Hi94}. Rather than abandon the RPWIA in favor of more 
sophisticated relativistic models, our approach has been to critically review the 
approximations and perform more refined calculations in order to reveal the limitations 
of the model.

In recent papers \cite{Hi94,Hi95} we calculated better effective masses  
and qualitatively analyzed the sensitivity of complete sets of quasielastic 
($\vec p,\vec p\ '$) and $(\vec p, \vec n)$ polarization transfer observables 
to medium effects, different forms of the $\pi$NN vertex, exchange contributions to the nucleon--nucleon 
(NN) amplitudes, and also spin--orbit distortions. We emphasized that the much--used 
SVPAT form \cite{Mc83a} is limited in that it does not properly
address the exchange behavior of the NN amplitudes in the nuclear medium, and does not properly 
distinguishing between pseudoscalar (PS) and pseudovector (PV) forms of the $\pi$NN vertex. Instead, we advocated the use of a meson--exchange model to 
explicitly include pions as one of the mediators of 
the NN interaction. For simplicity we considered the phenomenological Horowitz--Love--Franey 
(HLF) model \cite{Ho85} which parametrizes the relativistic SVPAT amplitudes as a sum of 
Yukawa--like meson exchanges in first Born approximation, and considers direct and exchange 
diagrams separately. Compared to the ($\vec p,\vec p\ '$) polarization transfer observables
the corresponding $(\vec p, \vec n)$ observables are generally
more sensitive to PS versus PV forms of the $\pi$NN vertex. Furthermore, most 
observables exhibited maximum sensitivity to nuclear medium effects at 
energies lower than 200 MeV. We also showed that, contrary to former expectations, exchange 
contributions are important in the entire 135 to 500 MeV range. 

Although our previous results stressed the potential value of quasielastic polarization transfer 
observables for studying nuclear medium effects, it failed to give an indication of the
experimental statistical uncertainty required for distinguishing between the various model predictions.  This was due to the fact that,
although the Fermi motion of the target nucleons yields NN scattering amplitudes
over a wide range of energies, in practice the
lack of published HLF parameter sets (at 135, 200, 300, 400 and 500 MeV) restricted us 
to consider only the parameter set closest to the incident laboratory kinetic energy for all effective laboratory kinetic energies. 
Hence, our results were merely qualitative and served only to give an initial ``feel'' for the 
sensitivities of observables to nuclear medium effects.

The aim of this paper, therefore, is to perform a similar, but quantitative study.
For a $^{40}$Ca target and a fixed momentum transfer of $1.97\ \mbox{fm}^{-1}$, we systematically investigate the sensitivity 
of complete sets of quasielastic ($\vec p,\vec p\ '$) and $(\vec p, \vec n)$ polarization transfer 
observables to medium effects, PS versus PV forms of the $\pi$NN vertex, and exchange 
contributions to the relativistic NN amplitudes. We have generated new HLF parameters 
(to be published in a future paper) between 80 and 195 MeV in 
5--MeV intervals, and utilized the recent Maxwell parametrization \cite{Ma96}, with both energy--dependent coupling 
constants and cutoff parameters between 200 and 500 MeV. 
The Fermi--averaging procedure, together with the availability of HLF parameters and reaction kinematics
of interest, restrict calculations to incident laboratory energies between 135 and 300 MeV.

Results are presented as ``difference '' graphs  in Figs.\ 1 -- 4: the solid and open circles respectively denote our calculated ($\vec p,\vec p\ '$) and 
$(\vec p, \vec n)$ values at the centroid of the quasielastic peak ($\omega \approx$ 80 MeV),
whereas the solid lines serve merely to guide the eye. 
We introduce the notation D$_{i' j}^{PS}(M^*)$ and D$_{i' j}^{{PV}}(M^*)$ to 
refer to polarization transfer observables calculated using respectively a PS and a PV
coupling for the ``pion'', both with the more refined effective masses $M_{SC}^{*}$ from Table II in Ref. \cite{Hi95}. The shaded 
areas accentuate differences between ($\vec p,\vec p\ '$) and ($\vec p,\vec n$) predictions. 

Per construction, the HLF-- and SVPAT--based D$_{i' j}^{PS}(M^*)$ observables
are identical. However, the Fermi--averaging procedure involves integrating over many 
amplitudes, and since the HLF parameter--fits are not perfect, slight differences on individual 
amplitudes could add constructively, thus translating to relatively large differences.
For polarization transfer observables, we found this theoretical uncertainty to be always 
smaller 0.04 and hence does not affect any of the conclusions drawn in 
this paper.

The sensitivity of polarization observables to PS versus PV forms of the
$\pi$NN vertex is denoted by $|D_{i' j}^{{PV}}(M^*)-D_{i' j}^{{PS}}(M^*)|$ in Fig.\ 1.
\begin{figure}
[t]
\centerline{\hspace{-2cm}\psfig{figure=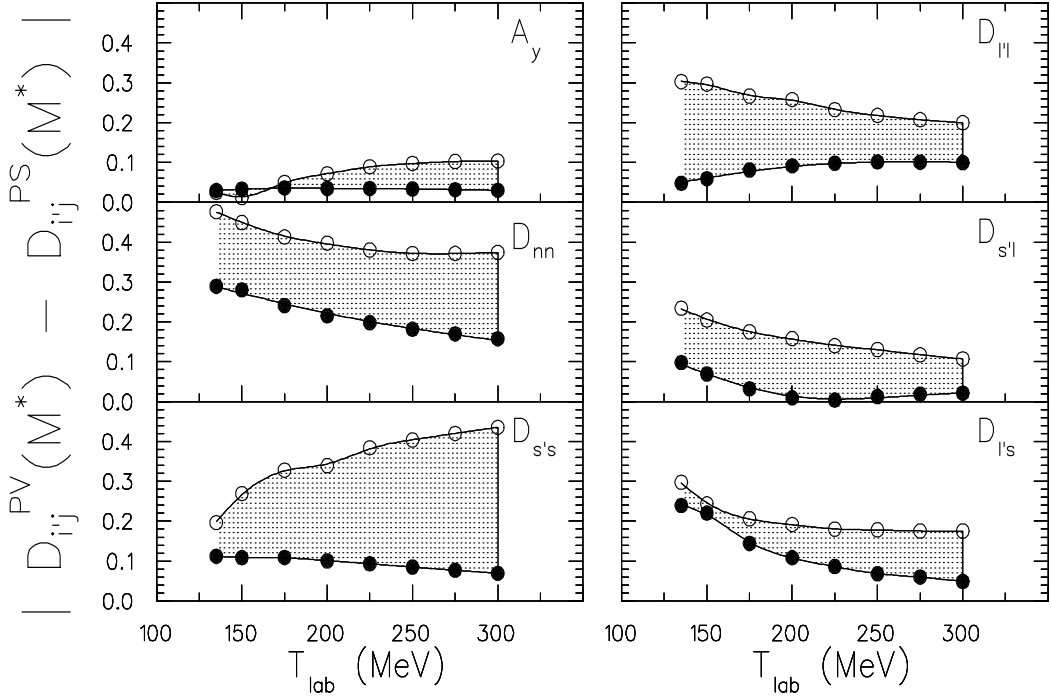,height=17cm,width=15cm}}
{\vspace{-7.5cm}\caption{The difference, $|D_{i' j}^{{PV}}(M^*) - 
D_{i' j}^{{PS}}(M^*)|$, for ($\vec p,\vec p \ '$) and 
($\vec p,\vec n $) polarization transfer observables  
observables D$_{i' j}$ calculated with a pseudovector (PV) and a 
pseudoscalar (PS) term in the NN interaction, respectively, as a function of 
laboratory energy and at the centroid of the quasielastic peak. 
Open circles represent ($\vec p,\vec n$) scattering, whereas
solid circles represent  ($\vec p,\vec p\ '$) scattering. 
The solid lines serve merely to guide the eye.}}
\label{fgpspvq}
\end{figure}

For ($\vec p,\vec n$) scattering, all the observables, except A$_{y}$, exhibit largest sensitivities 
to different pion couplings over the entire energy range. Generally, the sensitivities of the ($\vec p,\vec n$) 
polarization transfer observables completely overshadow the corresponding ($\vec p,\vec p\ '$) 
observables. Measurements of D$_{n n}$ for both ($\vec p,\vec n$) and ($\vec p,\vec p\ '$) scattering, 
particularly at low energies, would be extremely useful in shedding light on the preferred form of the 
$\pi$NN vertex.

Next, we choose a PS $\pi$NN vertex, and display the difference between effective--mass 
($M^{*}$) and free--mass ($M$)calculations in Fig.\ 2, denoted by $|D_{i' j}^{{PS}}(M^*)-D_{i' j}(M)|$.
\begin{figure}[t]
\centerline{\hspace{-2cm}\psfig{figure=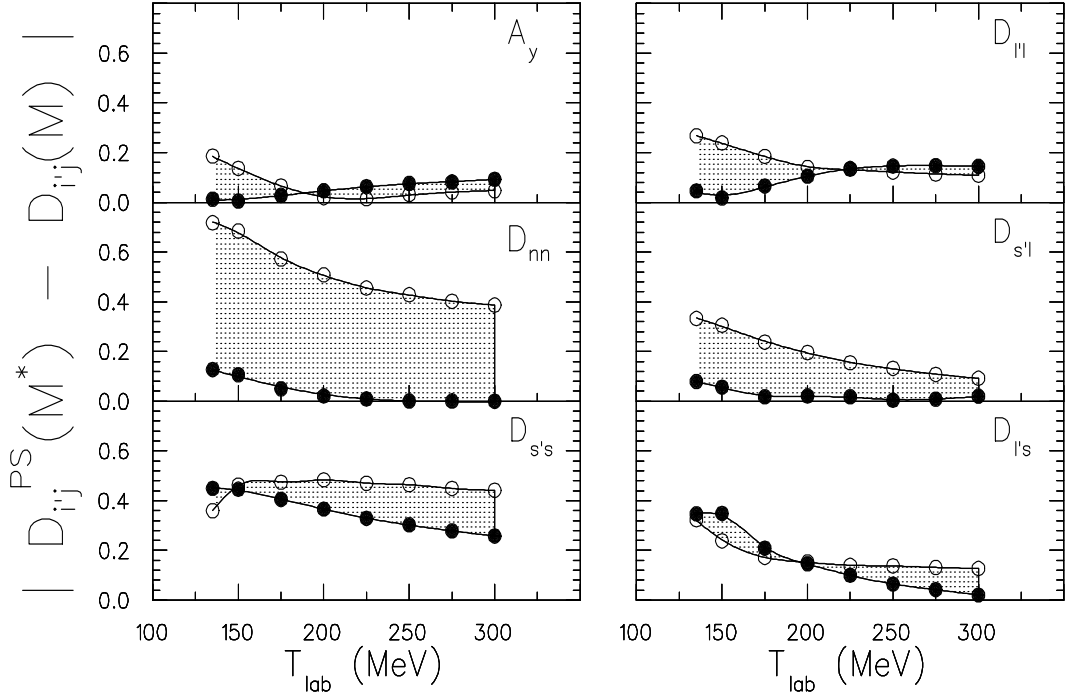,height=17cm,width=15cm}}
{\vspace{-7.5cm}\caption{Similar to Fig.\ 1, except that the values of 
$|D_{i' j}^{{PS}}(M^*)-D_{i' j}(M)|$ are plotted.}}
\label{fgmstpsq}
\end{figure}
These differences accentuate the importance of nuclear medium effects in polarization 
transfer observables. Compared to ($\vec p,\vec p \ '$) scattering, the ($\vec p,\vec n$) polarization 
transfer observables D$_{n n}$ and  D$_{s' \ell}$ are more sensitive to medium 
effects over the entire energy range. At higher energies for ($\vec p,\vec p \ '$) scattering, 
D$_{n n}$,  D$_{s' \ell}$ and D$_{\ell' s}$ observables are insensitive and 
correspond to free NN scattering. Note that D$_{n n}$ exhibits maximum 
and minimum sensitivity to medium effects for ($\vec p,\vec p\ '$) and ($\vec p,\vec n$) scattering 
respectively.

We now choose the PV form of the $\pi$NN vertex, and study the difference between effective--mass 
($M^{*}$) and free--mass ($M$)calculations. This is denoted by $|D_{i' j}^{{PV}}(M^*)-D_{i' j}(M)|$
in Fig.\ 3.
\begin{figure}[t]
\centerline{\hspace{-2cm}\psfig{figure=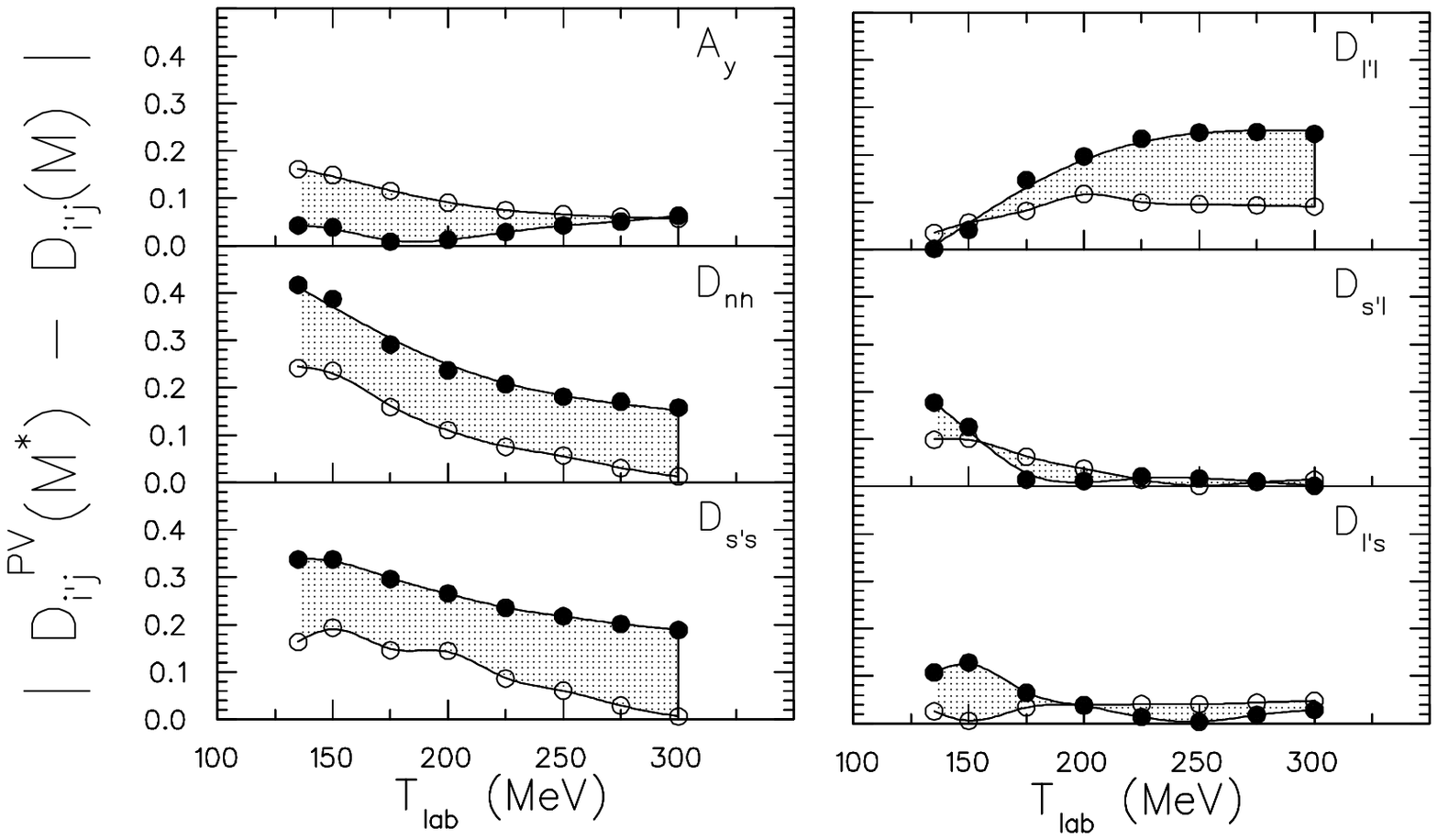,height=17cm,width=15cm}}
{\vspace{-7.5cm}\caption{Similar to Figs.\ 1 -- 2, except that
the values of $|D_{i' j}^{{PV}}(M^*)-D_{i' j}(M)|$ are plotted.}}
\label{fgmstarq}
\end{figure}
Compared to ($\vec p,\vec n$) scattering, the ($\vec p,\vec p \ '$) polarization 
transfer observables D$_{n n}$ and  D$_{s' s}$ are more sensitive to medium 
effects over the entire energy range. This is totally the opposite effect compared to the case 
with PS coupling. Hence, the effect of the nuclear medium depends critically on the type of pion coupling
for both ($\vec p,\vec n$) and ($\vec p,\vec p \ '$) scattering, particularly at low energies.
Comparison with experimental data (see later) will shed light on the type of coupling 
favored. At higher energies all the ($\vec p,\vec n$) observables are insensitive to medium 
effects and yield results similar to free NN scattering. 
Note the enhanced sensitivity of D$_{n n}$ and D$_{s' s}$ at low energies 
for both ($\vec p,\vec n$) and ($\vec p,\vec p \ '$) scattering.

Fig.\ 4 displays the sensitivity of polarization transfer observables to 
exchange contributions. For illustrative purposes, we choose the PV $\pi$NN vertex,
and plot the difference $|D_{i' j}^{PV}(M^*)_{Full} - D_{i' j}(M)_{Direct}|$
as a function of incident laboratory energy at the centroid of the quasielastic peak. The subscript 
``Full'' refers to the direct plus exchange amplitudes, whereas the subscript 
``Direct'' specifies amplitudes where exchange contributions are ignored.
\begin{figure}[t]
\centerline{\hspace{-2cm}\psfig{figure=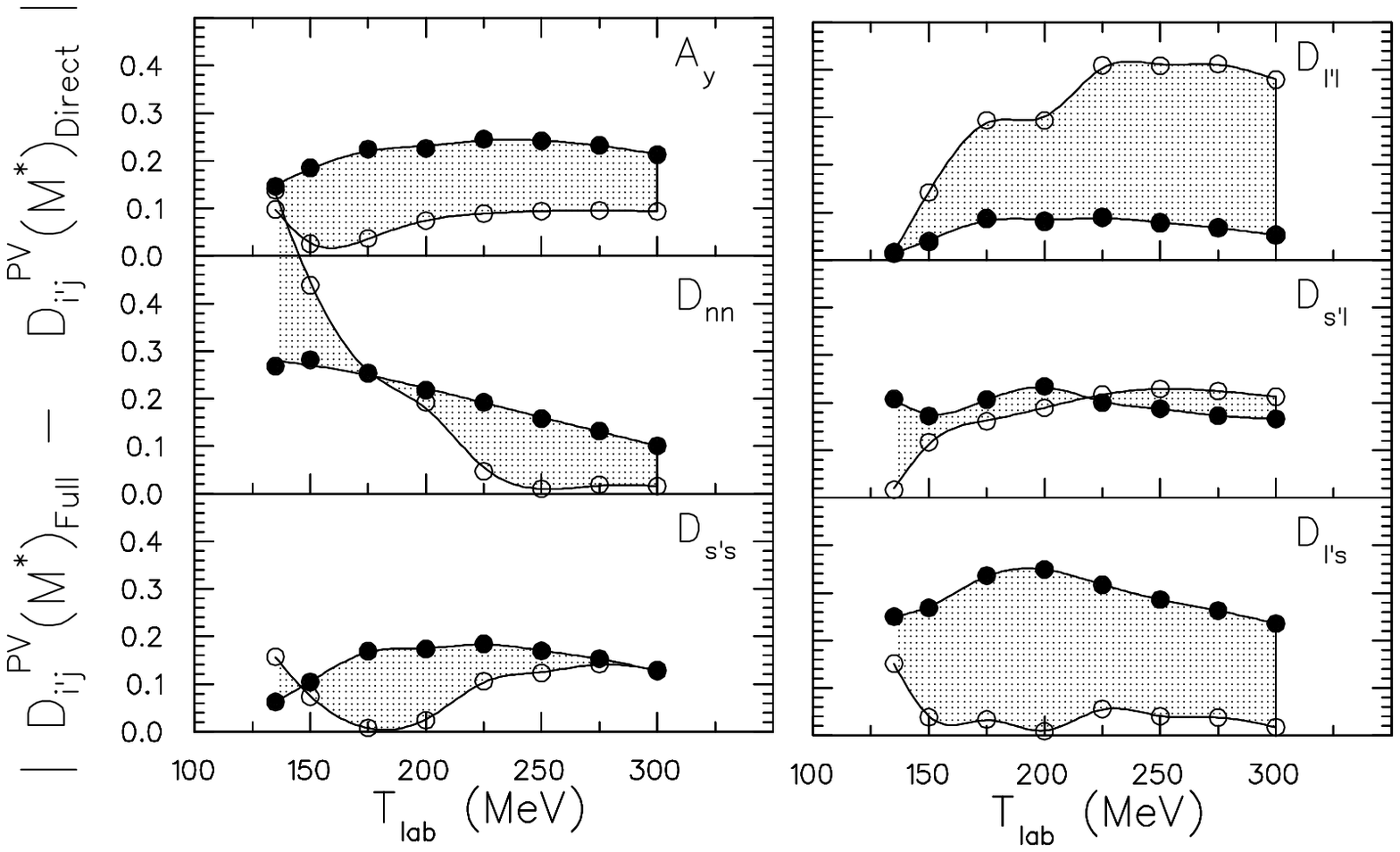,height=17cm,width=15cm}}
{\vspace{-7.5cm}\caption{Similar to Figs.\ 1 -- 3, except that
the values of $|D_{i' j}^{PV}(M^*)_{Full} - 
D_{i \ ' j}^{PV}(M^*)_{Direct}|$ are plotted.
The subscripts ``Direct'' and ``Full'' refer to calculations where the exchange 
terms have respectively been neglected and included respectively.}}
\label{fgexq}
\end{figure}
As in Ref. \cite{Hi95}, we see that for some polarization transfer observables the exchange contributions become 
more important at higher energies. In particular, for ($\vec p,\vec p\ '$) scattering, A$_{y}$ 
and D$_{\ell' s}$ are sensitive to exchange contributions over the 
entire energy range. Note the extreme sensitivity of D$_{n n}$ at low energies and D$_{\ell' \ell}$
at higher energies for ($\vec p,\vec n$) scattering. 

Finally, we compare HLF--model based RPWIA calculations to published 
experimental data. Results are displayed in Figs.\ 5 -- 6 and 
exclude spin--orbit distortions. The effect of spin--orbit distortions can be inferred from Ref. \cite{Hi94}.
The meaning of the various line--types is indicated in the figure captions.
The difference between the PS($M^{*}$)--SVPAT and PS($M^{*}$)--HLF calculations
gives an indication of the theoretical uncertainty attributed to the HLF model parameters:
this is typically much smaller than the statistical error bars.

Fig.\ 5 compares our calculations to $^{12}\mbox{C}(\vec p,\vec n)$ data 
at an incident energy of 186 MeV and momentum transfer 1.1 fm$^{-1}$ \cite{Wa94}. 
The centroid of the quasielastic peak is 
located at $\omega \approx$ 50 MeV.
The energy transfer $\omega$ includes the reaction Q--value of $-$18.6 MeV.
\begin{figure}[t]
\centerline{\psfig{figure=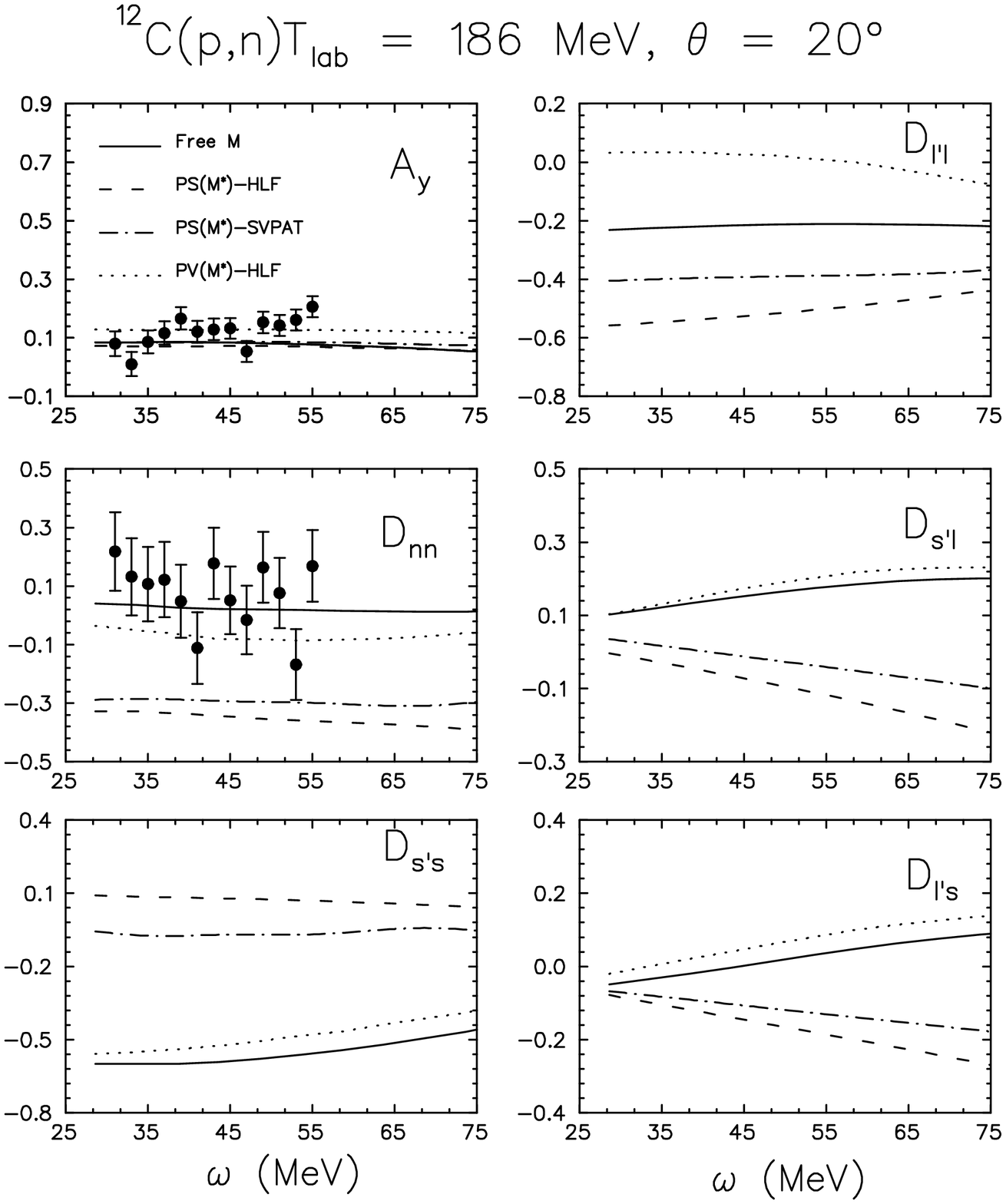,height=11cm,width=11cm}}
{\vspace{-0.5cm}\caption{Polarization transfer observables as a function of transferred energy $\omega$
over the quasielastic peak for $^{12}\mbox{C}(\vec p,\vec n)$ scattering at 186 MeV and 
$\theta_{\mbox{lab}}$=20${^\circ}$. The centroid of the 
quasielastic peak is situated at $\omega \approx$ 50 MeV.
Data are from Ref. [15]. The solid lines
indicate free--mass ($M$) calculations [Free $M$], dotted lines represent effective--mass ($M^{*}$)
PV calculations based on the HLF model [PV($M^{*}$)--HLF], dashed lines display effective--mass 
($M^{*}$) PS calculations based on the HLF--model [PS($M^{*}$)--HLF], and dashed-dotted lines show 
effective--mass ($M^{*}$) calculations based on a direct SVPAT parametrization of the 
Arndt phases [PS($M^{*}$)--SVPAT].}}
\label{fg186c20}
\end{figure}
D$_{n n}$ clearly favors a PV treatment of 
the $\pi$NN coupling, whereas A$_{y}$ fails to distinguish between PS and PV forms of 
the coupling. Note however, that both the free--mass 
and PV($M^{*}$)--HLF calculations describe the data 
equally well. The largest difference for the latter predictions occurs
for D$_{\ell' \ell}$; unfortunately the theoretical uncertainty
is also the largest for this observable. Hence, for all practical
purposes, the PV($M^{*}$) calculations are identical to the
free--mass calculations. It would be interesting to see whether this is verified experimentally
by comparing complete sets of $^{12}\mbox{C}(\vec p,\vec n)$ and 
$^{2}\mbox{H}(\vec p,\vec n)$ polarization transfer observables at 186 MeV.

Fig.\ 6 displays calculations for $^{12}\mbox{C}(\vec p,\vec p\ ')$
at an incident energy of 290 MeV and momentum transfer 1.97 fm$^{-1}$ \cite{Ch90}. 
The centroid of the quasielastic peak is 
located at $\omega \approx$ 80 MeV. 
\begin{figure}[t]
\centerline{\psfig{figure=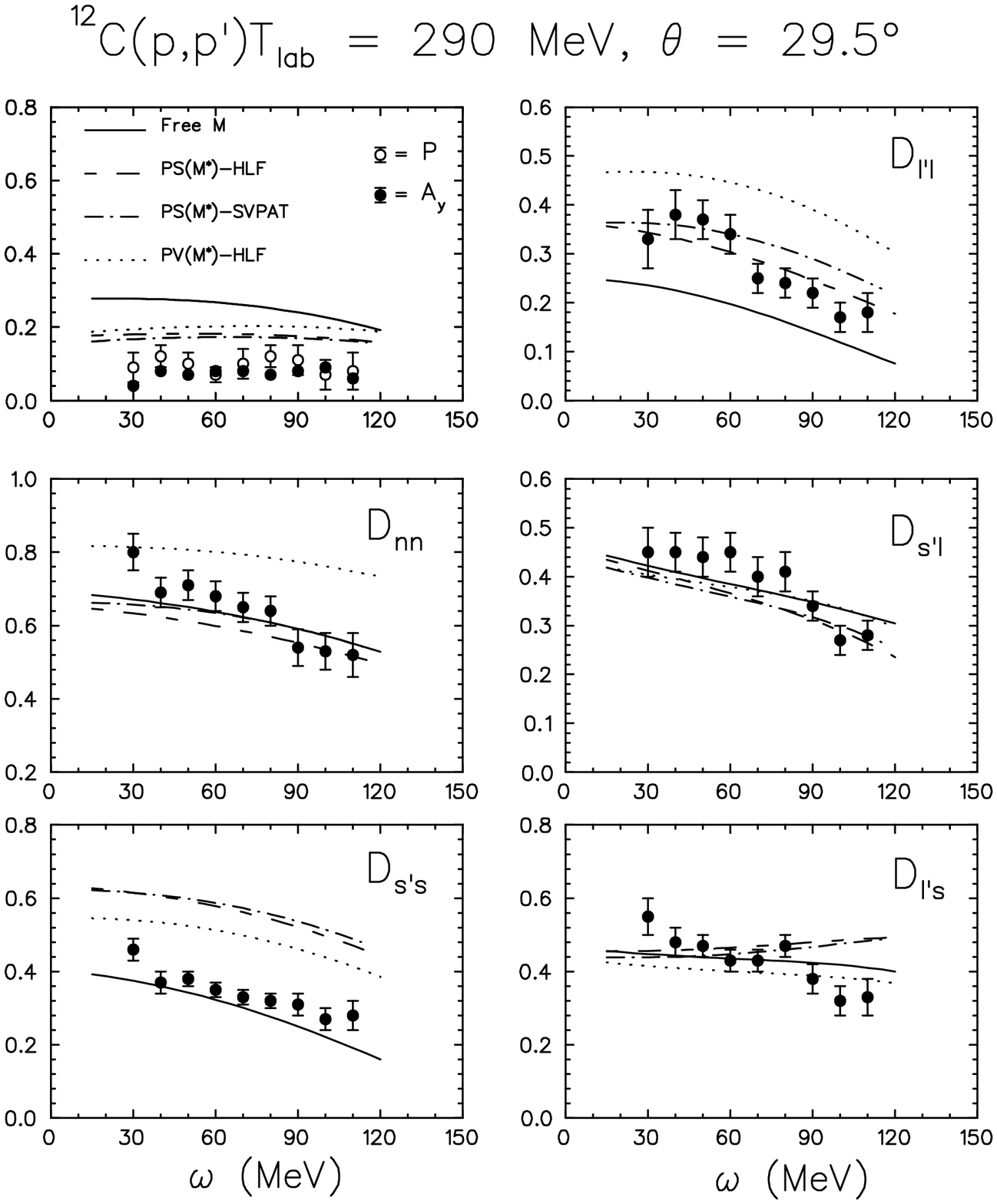,height=11cm,width=11cm}}
{\vspace{-0.5cm}\caption{Similar to Fig.\ 5, except that we now plot the 
polarization transfer observables for quasielastic $^{12}\mbox{C}(\vec p,\vec p\ ')$ scattering 
at 290 MeV and $\theta_{\mbox{lab}}$=29.5${^\circ}$. The centroid of the 
quasielastic peak is situated at $\omega \approx$ 80 MeV.
Data are from Ref. [16]. P and A$_y$ refer to the induced 
polarization and analyzing power respectively.}}
\label{fg290c30}
\end{figure}

We see that D$_{n n}$, D$_{s' s}$,
D$_{s' \ell}$ and D$_{\ell' s}$ correspond to the free--mass predictions.
Note that most of the observables favor a PS $\pi$NN vertex in contrast
to the PV form suggested by $(\vec p,\vec n)$ scattering.
None of the calculations predict A$_{y}$. However, the inclusion of spin--orbit distortion moves most of the 
medium--modified polarization transfer observables, including A$_{y}$, closer to the data. The effect 
of relativity is to quench A$_{y}$ for quasielastic ($\vec p,\vec p\ '$) scattering relative to the free 
mass values. To date all nonrelativistic models fail to predict 
this quenching effect. Note, however, that the celebrated ``relativistic signature'' \cite{Ho88} 
is much smaller than relativistic effects predicted for other polarization transfer observables at 
lower energies. For $(\vec p,\vec n)$ scattering with a PV $\pi$NN vertex, we predict a sizable medium effect 
on A$_{y}$ at $q = 1.97$ fm$^{-1}$. However at  $q = 1.1$ fm$^{-1}$
our calculations show no sensitivity to medium effects as is confirmed by the
limited IUCF data set \cite{Wa94}. Therefore it would be interesting to measure A$_{y}$
for a range of angles on a $^{12}$C target. Furthermore, we see that all calculations fail to describe the
D$_{s's}$ data.
As with the original RPWIA calculations, comparison with the small amount of 
available data still gives mixed but encouraging results. The $(\vec p,\vec p\ ')$ data 
favor a PS coupling for the pion, whereas the limited $(\vec p,\vec n)$ spin 
observable data suggest a PV form. The latter ambiguity can perhaps be 
attributed to the simple Born approximation embodied by the 
phenomenological HLF model. 
Furthermore, one should rather use a general Lorentz--invariant representation of the NN amplitudes 
as suggested by Tjon and Wallace \cite{Tj85}, instead of only the 5 SVPAT Fermi invariants.
Indeed we are currently investigating the former representation of the NN
scattering amplitudes for quasielastic proton--nucleus scattering.
A number of effects, which we have neglected, could also improve the
theoretical description of the data. For example, multiple scattering effects become sizable in heavy
nuclei and at large scattering angles \cite{Ho88}.
Furthermore, although signatures of low--lying collective states and giant resonances disappear
at the large excitation energies of interest, the nucleus continues to respond
collectively through the residual particle--hole interaction. This collectivity
manifests itself in gross features of the spectrum, such as shifts in the position
of the quasielastic peak and deviations of polarization transfer observables from the free 
values. Recently Horowitz and Piekarewicz \cite{Ho94} improved the simple Fermi--gas treatment 
of the nucleus by considering a relativistic random--phase approximation to the Walecka model.
Essentially this description takes into account the 
interactions between the nucleons in the medium at the mean--field level. 
These relativistic RPA correlations give a good description
of data and lead to an improvement over Fermi--gas predictions.
Furthermore, the effect of relativistic distortions on quasielastic polarization
transfer observables is still an open question. We are currently considering a full 
relativistic distorted wave description of quasielastic proton--nucleus scattering.

For both $(\vec p,\vec p\ ')$ and $(\vec p,\vec n)$ scattering, the number of observables 
that exhibit maximum sensitivity to nuclear medium effects, increase as the incident beam 
energy is lowered. In general, there is a lack of complete sets of 
published polarization data for quasielastic $(\vec p,\vec p\ ')$ and 
$(\vec p,\vec n)$ scattering at the intermediate energies of interest. 
In particular, at energies lower than 200 MeV there exists absolutely no 
complete published data set. Ideally one must measure complete sets of polarization 
transfer observables for both $(\vec p,\vec p\ ')$ and $(\vec p,\vec n)$ 
reactions for the same target, energy and momentum transfer.

\end{document}